# An Study of The Role of Software Project Manger in the Outcome of the Project

Israr Ali
Faculty of Engineering Sciences & Technology
Iqra University Main Campus
Karachi, Pakistan
Israr.ali@iqra.edu.pk

Aarij Mahmood Hussaan
Faculty of Engineering Sciences & Technology
Iqra University Main Campus
Karachi, Pakistan
aarijhussaan@iqra.edu.pk

Syed Hasan Adil
Faculty of Engineering Sciences & Technology
Iqra University Main Campus
Karachi, Pakistan
hasan.adil @iqra.edu.pk

*Abstract*—This paper describes an in depth analysis of successful and unsuccessful software Projects and the Role of Software Project Mangers in that success. One of the main reason in software project success is manager. Software houses are investing too much in this regard but the average ratio of software project failure is on the high side. Project managers experience, technical knowledge, and skills are not good enough for success in general. In this paper we have conducted a survey related to the approached used by different project managers, their methods and techniques, and the success ratio of their projects, and the steps they took during their projects. We will explore the core reasons of software project success and then will suggest key steps to be taken by the software project managers to deliver a successful software project.

## I. Introduction

The whole life cycle of a software project, from the start until the deployment has never been easy because it involves the optimized combination of the technical related skills and social skills of employees within the organization [1]. Therefore, all Lifecycle activities demands a professional skill set on the part of the software project manager. But, the problem is that even so many years of research and development, software development, management and its implementation are still very hard to control and results in failure [1]. There are studies highlighting the fact that software project managers must understand about how to manage, to deliver as a successful software project. A factor mentioned by many researchers and it is of course very important because it impacts software project success very highly, is the techniques used by software project manager and their skill set in the whole Lifecycle of the software project.

Studies were done and are available that discusses the required skill set descriptions of the software project managers. However, the problem is that no one could be found which helps in extracting an exact procedure for software project management skill set and their results which effects software project success. Therefore, this paper is an effort to find and propose a research model which will provide a way to find out the link between technical expertise set of managers, and the way they need to take towards project success. At this point we can say that our most hot research question which will be discussed in this paper is

**How is software project success derived by the expertise and skill set of software project managers?**

The secondary research question for this paper is:

**What Steps a software project manager need to take to guide the project towards success?**

This paper follows a normal survey based approach, for this a survey is conducted for explaining the research questions and to find the answers to these questions.

## II. Software Projects Success

Measuring the success of Software projects is complicated by varying definitions of success. Success can be judged by some important factors which includes deadlines, finance and features or services delivered [12].Some of the Success Factors associated with implementation projects having a very large scope includes achieving the complete support of employers [14]. Improving business processes before implementing developed solutions [9]. Satisfying an excellent involvement between users and the software organization is a success [17]. Implementation phase of software projects mostly results in surprises with respect to overall cost, because of the decentralized approach, user support, and software version control. [18]. Project management best practices include positive outside cooperation strategies, such as establishing a user satisfaction team, user feedback on a regular basis on the project, and customer learning level for implementation and installation [19]. Software project success depends on the way to which the software project accomplished its goals [7]. Software development best techniques which results in successful projects are well organized planning, change management with long term approach, business customer satisfaction, skills set of employees needed for the project, and led by an experienced manager.

## III. LITERATURE REVIEW

A review of several literatures helped us to understand the specific organizational requirements, individual persons requirements; software project managers skill sets requirements, software project implementation requirements and its success factors. The purpose of this review was to identify research that describes the skill set needed by software project managers. Lee [4] thinks that this kind of research essentially groups software project managers knowledge into three categories

- **Technology**
- **Business**
- **RelationShip**

There study suggest that the technical and business skill-sets of the mangers with the relationship of the manger with the customer and other important stakeholder is the key to success and are directly related to the software project further. Jef Coble [10] categorized the software systems into successful, challenged and failed. He describes different charts and his survey results related to software success. He delivers a chart display the percentage of software projects being regarded as successful, challenged or declared as Failure under the definition given below. Steven J. Lorenzet , Drew Procaccino and June M. Verner [9] tried to determine the process-related factors for software managers thinking of project success. Mark Keil [1] suggested an approach for identifying and removing the software project risks and how to handle and manage these risks. He indicated that mangers thought or thinking about the risk were higher for those projects over which they had little control. Mary Sumner [14] in his paper described the major failure reasons including. She describes some critical success factors as well as some major failure reasons. The most important factor she believes to break the project into smaller phases if possible. She further said that Phased-in approach in software project development is very much productive. Richard Berntsson-Svensson [13] research was to find out software project success factors in their own country software industries. As a matter of fact, their findings about the facts say yes to the results from other studies. These findings are very well-organized and with acceptable scope, good schedule estimations, well collected set of requirements, end-user involvement were the most essential factor contributing towards software project success, their suggestion was that adding extra team members to meet tough schedule estimations can be contributing to the project failure. Their results suggest that for different industries or organization the software project success factors are different. It simply means that software project manager needs to understand that they need to rethink their strategies for different type of organizations. Their conclusion was that the factors which are essential for software project success is not considering what type of system manager are going to develop and they also miss the concept of whether the system is being developed from zero or its a new system or whether the project is about the current features and scope of an existing system. Manzil-e-Maqsood and Talha Javed [11] have talked about the need now to educate the new students about the software project management so that when they will become the managers they will be ready to face these complicated issues. They discussed the flaws which are currently present in education strategy to teach software project management.

## IV. RESEARCH METHOD

The primary research question is to find out software project success in software development and the role of the software project manager in this success. The following research questions pop up in mind for our investigation:

- What are the factors on which the success or failure of software projects depends?
- How is software project can be successfull by using different kind of skills of software project managers?
- What Steps a software project manager need to take to guide the project towards success?

A detailed questionnaire was needed so it was designed to find out the answers of the research questions. Some initial studies were performed to optimize the questionnaire. The questionnaire which was finalized at the end has 10, close-ended questions. We have, some open-ended and also some weighted questions in the questionnaire. Ethical factors were kept in mind before sending or distributing the questionnaire. The first questionnaire contained questions related to personal information about managers for gathering background information about the project manager. After that the questionnaire focused on close-ended questions about the factors related to the success of a software project and the steps from managers and failure of software projects and the steps of managers. The questions focused on different factors and was answered using a well-defined grading criteria. This part also contained some useful open-ended questions to capture from managers their own thoughts and wordings. After that the questionnaire contained questions about software managers own methods and strategies, what methodology they use normally, what is their decision effect on success of the project and their skill level. This made it easy for our research to find importance of each factor and the relationship of those factors with software success.

*A. Data Collection*

Data is captured / collected from different software houses and organizations professionals. The participants in this study are software engineering and software project managers from the IT organization, educational universities, and software consulting agencies. A probability sampling strategy is used because it is easy to use and it provides a better result in the case of fewer records. Samples were distributed in different ways but collected in about many different ways; it was done by the support of our friends working in different organizations. The managers from companies were given the questionnaire by the friend we have in that company, or the friend our friend has in that company. This study used many forms to collect data but most important were interviewed and questionnaires emailed to the managers. Those questionnaires which were emailed actually processed early because of the immediate response. But still emailed ones had problems because of which another approach that is interviews with the managers by us or by any of our friends were used to increase the authority of result and the count of result as well as the size of the data collected. After doing all these stuffs and collecting all the data we divided the data into success and failure categories and concluded our results. The Survey was distributed the leading software houses from all over the Pakistan. This survey was intended for the software project managers and although it was tough to collect feedback from all these managers but eventually it paid as results matters most. The project managers who shared their thoughts on this research belongs to leading software houses including Kalsoft , Avanza Solutions, Axact, Folio3 , Pixsense, Trg Tech, Corrtec, Mazik, Tenpearls ,eDev Technolgies, Visual Soft, Mksoft, SoftTech , Algorithm, Inbox, Etilize, Matrix , TPS, ANOVA

## V. RESULTS

Following are the results we gathered from different software houses.

### A. Successfully Implemented Software Projects

In analyzing research questions, this section compiles the success factors for software project managers understandings. After detailed considerations on the literature, we choose some factors specifically related to project success, as given in Figure 1. Mangers answered each factor and told us which factors most strongly affected the projects that they participated in. Factors for the managers and their responses suggested the major factors to consider in planning towards a successful software project. The Results from these questions suggest that for a successful software project the manager needs to hire people having the domain knowledge of the project, Top management involvement must be made necessary, The requirement should always be delivered to the team very clearly, business expert accessibility is a major part in a software project, the changes in the project should be an expected phenomena and should be handled smartly and there should be some pressure on the development team to deliver milestones.

### B. Software Solutions Not Successfully Implemented

The Survey included some question related to the software solution not successfully implemented. Some Factors related to the Failure of a software solution were filled out by different managers and their response is given in Figure 2. The Results suggest that the major factors which should be avoided by a manager is to not hire project persons out of domain expertise of a project, customer satisfaction and respect must not be negotiated within the team, there should be smart handling of expectations and organizational difficulties. There should not be long working hour for the team in which they feel unnecessary pressure and team persons should be kept intact.

### C. Successful Project Manager Contributions

The Survey included a Question for the software project managers related to the contribution in the different phases of the project. The purpose of this question was to find out the contribution of successful manager as well as the contribution of unsuccessful in every kind of project life cycle phase. The Figure given below indicates the project managers contributions per project phases that have excellent project success rates.

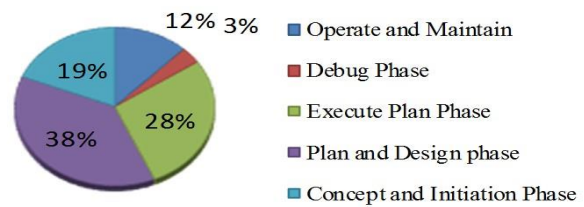

Fig. 3. Contribution of Successful Project Manager Decisions per Project Phase

The results suggest that the largest contribution from mangers are in Plan and Design Phase and then the Execute plan phase, remember it is for the project managers who have a larger success rate. Therefore it can be judged that to be successful the mangers need to emphasize on planning and design phase.

### D. Unsuccessful Project Manager Contributions

The Figure given below indicates the project managers contributions that have not very good project success rate.
The results clearly suggest that the project manager contribution to the concept and initialization phase and plan and design phase is less as compared to contributions of successful managers, suggesting that these managers need to focus on the initial phases of a project life cycle. An interesting observation is that unsuccessful mangers have larger contributed in executing plan phase.

Another important question in the survey was the Project Manager Decision as Project Success Factor and the results of data from a survey is given below in a chart.

Results suggest that the project managers are so what puzzled that their decision would make a difference in the success of the project or failure. A majority believes that their decision will definitely make a difference; same numbers of mangers were little confused about this question.

|   | Factor | SA | A | N | D | SD |
|---|---|---|---|---|---|---|
| 1 | Domain knowledge, expertise of project persons was Excellent. | 70 | - | 30 | - | - |
| 2 | Top management commitment to the project was Excellent. | 100 | - | - | - | - |
| 3 | Customer/user and team had good relationship. | - | 60 | 40 | - | - |
| 4 | Team Understood the requirements very well. | - | 80 | - | 20 | - |
| 5 | Code Auditing was an essential part. | - | 50 | 30 | 20 | - |
| 6 | End user expectations were Managed Successfully. | 100 | - | - | - | - |
| 7 | Change of Requirements were expected and managed. | 60 | - | - | 40 | - |
| 8 | Team was Included in decision making. | - | 30 | - | 70 | - |
| 9 | Sufficient/appropriate staffing as per project requirement. | 100 | - | - | - | - |
| 10 | Code versioning was an important factor. | - | 40 | 50 | 10 | - |
| 11 | Access to business / technical experts was good enough. | 100 | - | - | - | - |
| 12 | Bug Tracking and solving was standardized. | - | 80 | 20 | - | - |
| 13 | Team was able to negotiate changes. | 100 | - | - | - | - |
| 14 | Financial resources were good enough. | - | 20 | 50 | 30 | - |
| 15 | Urgency in development was an important factor. | 100 | - | -- | - | - |

Fig. 1. The Factors for Software Success provided by Managers

|   | Factor | SA | A | N | D | SD |
|---|---|---|---|---|---|---|
| 1 | The Budget allocated. | - | 40 | - | 60 | - |
| 2 | Domain knowledge, expertise of project personals. | - | 100 | - | - | - |
| 3 | Time scheduled for the project. | - | 40 | - | 60 | - |
| 4 | The reason for failure was customer. | 100 | - | - | - | - |
| 5 | Absence of a long term vision and requirements engineering. | - | 40 | 30 | 30 | - |
| 6 | Not possible expectations due to estimated problems and politics. | - | 80 | 20 | - | - |
| 7 | Lack of decomposition in team hierarchy | - | 20 | - | 80 | - |
| 8 | bad staffing policies and conflict with team in this manner | - | - | - | 100 | - |
| 9 | Minimum of all stakeholder feedback and involvement | - | 100 | - | - | - |
| 10 | Minimum of strategic mind and CEO support | - | 20 | - | 80 | - |
| 11 | Software project manager wanted long working hours | 100 | - | - | - | - |
| 12 | Adding extra team members to meet schedule | - | 50 | - | 50 | - |

Fig. 2. The Factors for Software Failure Reasons provided by manager

VI. CONCLUSION

In this paper we found the major reasons for the success of a software project and failure. After that a quick literature overview was given in which different statistical data were presented by using the different quotes of researchers. We explored that Business skill set, technical skill set and relationship strategies used by software project managers have a strong impact on software project success directly, therefor must be dealt with extra care. A Survey for manger was designed for this study and the results of survey suggested some important factors to the mangers about the success and failure of the project. These factors included better team building, excellent requirement gathering, good customer / user relationship, smart change management and good leadership. We explored what kind of contribution mangers put in when they are successful and when they are unsuccessful, and also what are their thoughts about their decisions making a difference in the success rate of the software project.

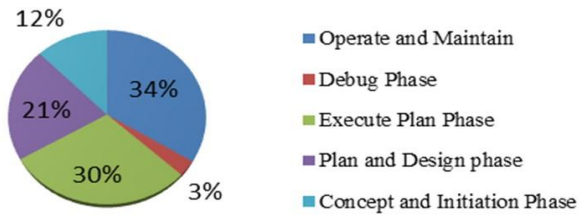

Fig. 4. Contribution of Unsuccessful Project Manager Decisions per Project Phase

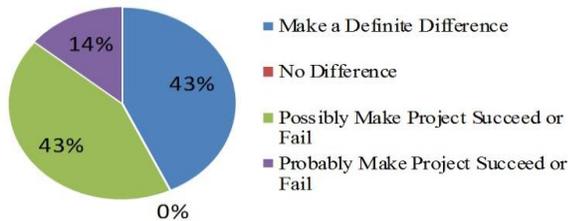

Fig. 5. Project Manager Decision Making as Project Success Factor

## VII. FUTURE DIRECTIONS

Some more information is required to get more desirable results. A larger data from survey and accurate statistical calculations will produce a large accuracy. In addition, studying different projects from same manager or similar project from different manager will boost the accuracy of the suggested work. Future studies will involve the strategies to find out of software success factors for various industries.

## VIII. ACKNOWLEDGEMENT

We would like to thank all of the software project managers, team leads and their companies/Organization who has helped in making the research possible for us in this paper.